\documentclass[%
reprint,
superscriptaddress,
amsmath,amssymb,
aps,
pra,
]{revtex4-1}

\usepackage{graphicx}
\usepackage{subfigure}
\usepackage{dcolumn}
\usepackage{bm}
\usepackage{braket}
\usepackage{xcolor}
\usepackage[colorlinks, linkcolor=red, anchorcolor=blue,urlcolor=blue, citecolor=blue]{hyperref}
\usepackage{amsmath}
\usepackage{academicons}

\begin{document}

	\title{Optimal control and ultimate bounds of 1:2 nonlinear quantum systems}
	
\author{Jing-jun Zhu
\href{https://orcid.org/0000-0002-4277-1730}{\includegraphics[scale=0.05]{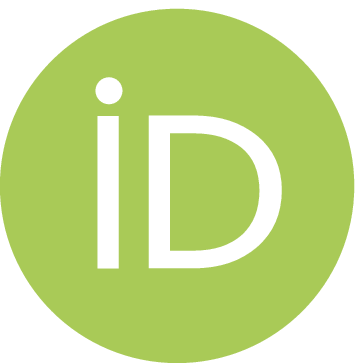}}}
\affiliation{Laboratoire Interdisciplinaire Carnot de Bourgogne, CNRS UMR 6303, Universit\'e de Bourgogne,
	BP 47870, 21078 Dijon, France}
\affiliation{International Center of Quantum Artificial Intelligence for Science and Technology (QuArtist) and Department of Physics, Shanghai University, Shanghai 200444, China}
\author{Kaipeng Liu
\href{https://orcid.org/000-0002-5642-8458}{\includegraphics[scale=0.05]{ORCIDiD.eps}}}
\affiliation{International Center of Quantum Artificial Intelligence for Science and Technology (QuArtist) and Department of Physics, Shanghai University, Shanghai 200444, China}

\author{Xi Chen
\href{https://orcid.org/0000-0003-4221-4288}{\includegraphics[scale=0.05]{ORCIDiD.eps}}}
\affiliation{Department of Physical Chemistry, University of the Basque Country UPV/EHU, Apartado 644, 48080 Bilbao, Spain}
\affiliation{EHU Quantum Center, University of the Basque Country UPV/EHU, Barrio Sarriena, s/n, 48940 Leioa, Spain}

\author{St\'ephane Gu\'erin \href{https://orcid.org/0000-0002-9826-95988}{\includegraphics[scale=0.05]{ORCIDiD.eps}}}
\email{sguerin@u-bourgogne.fr}
\affiliation{Laboratoire Interdisciplinaire Carnot de Bourgogne, CNRS UMR 6303, Universit\'e de Bourgogne,
	BP 47870, 21078 Dijon, France}
\date{\today}
	\begin{abstract}	
Using optimal control, we establish and link the ultimate bounds in time (referred to as quantum speed limit) and energy of two- and three-level quantum nonlinear systems which feature 1:2 resonance. Despite the unreachable complete inversion, by using the Pontryagin maximum principle, we determine the optimal time, pulse area, or energy, for a given arbitrary accuracy. We show that the third-order Kerr terms can be absorbed in the detuning in order to lock the dynamics to the resonance.
In the two-level problem, we determine the non-linear counterpart of the optimal $\pi$-pulse inversion for a given accuracy.
In the three-level problem, we obtain an intuitive pulse sequence similar to the linear counterpart but with different shapes. We prove the (slow) logarithmic increasing of the optimal time as a function of the accuracy. 
	\end{abstract}

\maketitle
	
\section{Introduction}
The accurate control of quantum dynamics is at the core of the quantum world. Quantum control protocols have been developed in order to design specific shaped pulses  including composite \cite{CPlevitt,Wimperis,CPprl106,Jones,CPprl129,CPnjp24}, adiabatic \cite{STIRAP,STIRAP2,UltrafastSTIRAP}, shortcut to adiabatic \cite{STA,robustNJP,STA2,pra103} and single-shot shaped pulse \cite{SSprl111,Hybrid,Laforgue} techniques.
However, these protocols, even when accelerated compared to standard adiabatic passage, 
do not specifically control the time of operation, 
which can lead to severe obstructions to experimental implementation. 
In this context, optimal control theory (OCT)  \cite{OCT} has emerged as a powerful tool to mitigate intensities of pulses 
allowing one to attain the ultimate time bound in the system, which is also interpreted as quantum speed limit (QSL) \cite{prl103_Tomaso,prl111,Frey,pra98}. 
Besides numerical implementation of OCT, such as monotonically convergent iteration algorithm \cite{OPalgorRuths134,OPalgorzhu108,OPalgorzhu110,nonOppra78}, global Krotov method \cite{Krotov}, gradient ascent algorithms (GRAPE) \cite{GRAPE}, 
one can highlight
Pontryagin maximum principle (PMP) \cite{PMP,Boscain,Extended,PMPprx}, which, transforming the initial infinite-dimension control problem into a finite dimension problem, allows analytic derivation of the optimal controls (typically with respect to time or energy). One can also mention recent geometric approaches \cite{Barnes,Dridi,Dridi2} treating simultaneously robust and optimal control. 

The extension of quantum control techniques to non-linear quantum systems relevant to describe BEC, e.g. when one considers the conversion from atomic to molecular BEC leading to a so-called 1:2 Fermi resonance \cite{pra65_Drummond}, is a non-trivial issue. The system has to be reinterpreted and analyzed with tools from classical mechanics, where the concept of integrability,  without counterpart in the standard linear quantum physics, plays an important role \cite{classical_H,Henrard}. When the system is integrable, adiabatic passage techniques can be formulated with trajectories formed by the instantaneous (stable) elliptic fixed points defined at each value of the adiabatic parameters and continuously connected to the initial condition. Obstructions to classical adiabatic passage are given by the crossing of a separatrix \cite{prl99_Itin,pra_SG,epl_stephane}. In addition, for a two-level problem with a 1:2 resonance, the north pole of the generalized Bloch sphere (associated to the upper state and thus corresponding to a complete population transfer from the ground state) is unstable since it is associated to an hyperbolic fixed point in the classical phase space representation \cite{pra102_Jingjun}. This prevents adiabatic passage to be robust when it approaches the north pole. The system is not controllable at this point in the sense that the nonlinearity prevents reaching the upper state exactly \cite{pra_SG}. However, one can approach it as closely as required, and inverse-engineering techniques \cite{prl119_Stephane,STAOPnon} have been developed for that purpose.

Ultimate bounds, e.g. quantum speed limit \cite{prl111}, can be defined via the minimization of a given cost (such as time, pulse area or energy) determined from optimal control; their extension to nonlinear systems is an open question. The purpose of this work is to establish and link these ultimate bounds in terms of time and energy using optimal control. We present a complete study of optimal control via the PMP for the two- and three-level systems featuring a 1:2 resonance, considering the cost as time or energy. Since the complete inversion from the ground state is unreachable, we define the target with a given (arbitrary) accuracy. This work completes the analysis made in \cite{pra94xc} for the nonlinear two-level system. 
Among the results, we extend the notion of (optimal) $\pi$-pulse \cite{Boscain} to nonlinear systems (for a given accuracy) and show the asymptotic logarithmic increasing of the optimal time as a function of the accuracy instead of Rabi oscillations.
We also show the similarity with the linear case: the nonlinear dynamics is shown to be identical for time or energy optimum with a constant pulse (or constant generalized pulse in the case of the three-level system).

Section \ref{model-1} and \ref{model} are devoted to two- and three-states problems, respectively. We conclude in Section \ref{conclusion}.

\section{1:2 nonlinear two-level model}
\label{model-1}
\subsection{The model}
The two-level model including second-order (with a 1:2 resonance) and third-order Kerr nonlinearities is characterized by following equations of motion  \cite{pra_SG}
\begin{subequations}
\label{motiontwolevel}
\begin{align}
	&i\dot{\psi}_{1}=\left[-\frac{\Delta}{3}+\Lambda_{11}|\psi_{1}|^{2}+\Lambda_{12}|\psi_{2}|^{2}\right]\psi_{1}+\frac{\Omega}{\sqrt{2}}{\psi}^{*}_{1}\psi_{2},\\
	&i\dot{\psi}_{2}=\left[\frac{\Delta}{3}+\Lambda_{21}|\psi_{1}|^{2}+\Lambda_{22}|\psi_{2}|^{2}\right]\psi_{2}+\frac{\Omega}{2\sqrt{2}}{\psi}^{2}_{1},
\end{align}
\end{subequations}
where $\psi_{1}$ and $\psi_{2}$ are the state probability amplitude, satisfying $|\psi_{1}|^{2}+2|\psi_{2}|^{2}=1$ that can vary in the respective ranges $|\psi_{1}|^{2}\in[0,1]$, $|\psi_{2}|^{2}\in[0,1/2]$.  The controls are time-dependent: $\Delta\equiv\Delta(t)$ and $\Omega\equiv\Omega(t)$, representing the detuning and Rabi-frequency, respectively. Here $\Lambda_{ij} (i,j=1,2)$ denote the third-order nonlinearities (in units of angular frequency) and $\Lambda_{21}=\Lambda_{12}$. Second-order nonlinearities appear in the form of the coupling. 
We will use units such that $\hbar=1$.

We can describe the dynamics on a generalized nonlinear Bloch sphere (see, e.g., \cite{pra102_Jingjun,Efstathiou}) by introducing the nonlinear coherences and the population inversion, respectively:
\begin{subequations}
\begin{align}
\eta_{1}&=\sqrt{2}\,\text{Re}\bigl(\psi^{2}_{1}\bar{\psi}_{2}\bigr),\quad \eta_{2}=\sqrt{2}\,\text{Im}\bigl(\psi^{2}_{1}\bar{\psi_{2}}\bigr),\\
\eta_{3}&=|\psi_{2}|^{2}-\frac{1}{2}|\psi_{1}|^{2}, \quad \eta_{3}\in\Bigl[-\frac{1}{2},\frac{1}{2}\Bigr],
\end{align}
\end{subequations}
 leading to 
 \begin{align}
 |\psi_{1}|^{2}=\frac{1}{2}(1-2\eta_{3}),\quad |\psi_{2}|^{2}=\frac{1}{4}(1+2\eta_{3}).
 \end{align}
 The generalized 1:2 nonlinear Bloch sphere is characterized by the following surface equation
\begin{align}
	\label{surface}
	\eta^{2}_{1}+\eta^{2}_{2}-\Bigl(\frac{1}{2}-\eta_{3}\Bigr)^{2}\Bigl(\frac{1}{2}+\eta_{3}\Bigr)=0.
\end{align}
The south and north poles correspond to $|\psi_{1}|^{2}=1$, $|\psi_{2}|^{2}=0$, i.e. $\eta_3=-1/2$ and $|\psi_{1}|^{2}=0$, $|\psi_{2}|^{2}=1/2$, i.e. $\eta_3=1/2$, respectively. It has been proved that $\eta_3=1/2$ is an unreachable (unstable) target with or without Kerr terms  in \cite{pra_SG,epl_stephane}. Using Eq. \eqref{motiontwolevel}, we have
\begin{subequations}
\label{etad}
\begin{align}
	\label{eta1}
		&\dot{\eta}_{1}=\bigl(-\Delta+\Lambda_a -2\Lambda_s|\psi_{2}|^2 \bigr)\eta_{2},\\
		\label{eta2}
		&\dot{\eta}_{2}=\frac{\Omega}{2}\Bigl(3\eta^{2}_{3}-\eta_{3}-\frac{1}{4}\Bigr)+\bigl(\Delta- \Lambda_a +2 \Lambda_s |\psi_{2}|^2\bigr)\eta_{1},\\
		&\dot{\eta}_{3}=\Omega\eta_{2}
		\label{eta3}
\end{align}
\end{subequations}
with the effective third-order nonlinearities 
 \begin{equation}
\Lambda_s= 2\Lambda_{11} +\Lambda_{22}/2 -2\Lambda_{12}, \quad \Lambda_a=2\Lambda_{11}-\Lambda_{21}.
 \end{equation} 
 It can be seen that $\Lambda_a$ can be trivially compensated by a static shift of the detuning, while the term proportional to $\Lambda_s$ depends on the dynamical variable $|\psi_{2}|^2$. However it has been shown in \cite{prl119_Stephane} that one can lock the resonance using the freedom in the choice of the time-dependance of $\Delta$ by incorporating the term $2 \Lambda_s |\psi_{2}|^2$.
Hence, we define the effective detuning (which includes a change of sign of $\Delta$ for convenience):
 \begin{equation}
 \label{DefTD}
 \tilde{\Delta}=-\Delta+\Lambda_a -2\Lambda_s|\psi_{2}|^2=-\Delta+\Lambda_a -\Lambda_s \Bigl(\frac{1}{2}+\eta_{3}\Bigr),
 \end{equation} 
 such that the set \eqref{etad} of differential equations only features the second-oder nonlinearity:
 \begin{subequations}
\label{etad_}
\begin{align}
	\label{eta1_}
		&\dot{\eta}_{1}=\tilde\Delta\eta_{2},\\
		\label{eta2_}
		&\dot{\eta}_{2}=\frac{\Omega}{2}\Bigl(3\eta^{2}_{3}-\eta_{3}-\frac{1}{4}\Bigr)-\tilde\Delta\eta_{1},\\
		&\dot{\eta}_{3}=\Omega\eta_{2}.
		\label{eta3_}
\end{align}
\end{subequations}
This generalizes the standard Bloch equations to the 1:2 nonlinear system with the detuning  $\tilde{\Delta}$ \eqref{DefTD} featuring an explicit (linear) dependance on $\eta_3$ via the effective third-order nonlinear term $\Lambda_s$.
To determine the expression of optimal trajectories in this system, we apply the PMP taking for the cost the time (time optimal control) or the energy (energy optimal control). The 1:2 nonlinear two-state time optimal control has been solved in  \cite{pra94xc}, and we closely follow its derivation in order to apply it for the energy optimal control and in next section for the three-level problem.

To achieve a given transfer from a population $\eta_{\text{3i}}\equiv\eta_3(t_i)$ at the initial time $t_i$, we define the targeted final population $\eta_{\text{3f}}\equiv\eta_3(t_f)=\frac{1}{2}-\epsilon$ at the final time $t_f$, i.e. $|\psi_{2}(t_f)|^{2}=\frac{1}{2}(1-\epsilon)$. When one targets the upper state, the deviation $\epsilon$ will be taken small but different from zero since the second-order nonlinearity prevents to reach it exactly. 

\subsection{Time and area optimal control}
Various situations have been considered in \cite{pra94xc}. Here, we particularly focus on the (almost) complete transfer from the ground state as a function of $\epsilon$, and will show that the minimum time (or pulse area) increases in a (slow) logarithmic way with respect to small deviations $\epsilon$, or reciprocally that the deviation $\epsilon$ decreases exponentially with respect to the minimum time (or pulse area).

We consider the time minimizing functional 
\begin{align}
\label{MinTimeCost}
	J=\int_{t_i}^{t_f}dt.
\end{align}
The corresponding control (or pseudo) Hamiltonian from the set \eqref{etad_} of differential equations is (where we have added a constant $p_{0}$):
\begin{equation}
\label{Def_Hc}
	\begin{split}
		&h_{c}=\tilde{\Delta}(\lambda_{1}\eta_{2}-\lambda_{2}\eta_{1})+\Omega\Bigl[\frac{\lambda_{2}}{2}\Bigl(3\eta^{2}_{3}-\eta_{3}-\frac{1}{4}\Bigr)+\lambda_{3}\eta_{2}\Bigr]
	\end{split}
\end{equation}
with Hamiltonian's equation for the (dimensionless) costate $\Lambda=[\lambda_1,\lambda_2,\lambda_3]^{\mathsf{T}}$ gathering the conjugate momenta of $\eta_1$, $\eta_2$, and $\eta_3$, respectively:
\begin{subequations}
\begin{align}
	&\dot{\lambda}_{1}=-\frac{\partial h_{c}}{\partial\eta_{1}}=\lambda_{2}\tilde{\Delta},\\
	&\dot{\lambda}_{2}=-\frac{\partial h_{c}}{\partial\eta_{2}}=-\lambda_{1}\tilde{\Delta}-\lambda_{3}\Omega,\\
	&\dot{\lambda}_{3}=-\frac{\partial h_{c}}{\partial\eta_{3}}=-\frac{ \lambda_{2}}{2}\Omega(6\eta_{3}-1)-\frac{\partial\tilde{\Delta}}{\partial\eta_{3}}(\lambda_{1}\eta_{2}-\lambda_{2}\eta_{1}).
	\label{dlamb3}
\end{align}
\end{subequations}
In order to prevent arbitrary large field amplitude detrimental for experimental implementation, we impose a boundary on the field $\Omega\le\Omega_{0}$ as a constraint. 
The maximization of $h_c$  according to the PMP corresponds thus to the necessary condition
\begin{align}
	\frac{\partial h_{c}}{\partial\tilde{\Delta}}&=0,
\end{align}
i. e. 
\begin{equation}
	\lambda_{1}\eta_{2}-\lambda_{2}\eta_{1}=0.
	\label{p1}
\end{equation}
This leads to the shape of the external field $\Omega$:
\begin{equation}
	\Omega=\frac{2}{\lambda_{2}\left(3\eta^{2}_{3}-\eta_{3}-\frac{1}{4}\right)+2\lambda_{3}\eta_{2}}
	\label{Omega}
\end{equation}
from (\ref{Def_Hc},\ref{p1}) and the fact that the system \eqref{Def_Hc} is autonomous, i.e. $h_{c}=const.$ In Eq. \eqref{Omega}, we have renormalized $\lambda_2/h_c\to\lambda_2$ and $\lambda_3/h_c\to\lambda_3$ without loss of generality. 
Differentiating Eqs. \eqref{p1} and \eqref{Omega} gives
\begin{align}
	\label{p3}
	&\frac{\lambda_{1}}{2}\left(3\eta^{2}_{3}-\eta_{3}-\frac{1}{4}\right)+\lambda_{3}\eta_{1}=0,\\
	&\dot{\Omega}=0,
\end{align}
from which we conclude that $\Omega$ is constant, taken at its maximum $\Omega=\Omega_0$. 
Multiplying Eq. \eqref{Omega} by $\eta_{1}$ and Eq. \eqref{p3} by $\eta_{2}$ also using \eqref{p1}, we obtain a linear system of equations for the variables $\lambda_{2}$ and $\lambda_{3}$:
\begin{subequations}
	\label{p4}
	\begin{align}
		&\lambda_{2} \frac{\eta_{1}}{2}\Bigl(3\eta^{2}_{3}-\eta_{3}-\frac{1}{4}\Bigr)+\lambda_{3}\eta_{1}\eta_{2}=\frac{\eta_{1}}{\Omega},\\
		&\lambda_{2}  \frac{\eta_{1}}{2}\Bigl(3\eta^{2}_{3}-\eta_{3}-\frac{1}{4}\Bigr)+\lambda_{3}\eta_{1}\eta_{2}=0
	\end{align}
\end{subequations}
of determinant zero, which can give a solution when the inhomogeneous terms are zero, i.e. $\eta_1=0$. This implies $\lambda_{1}=0$ from Eq. \eqref{p3} (for a non-constant $\eta_3$) and $\dot{\eta}_{1}=0$ in Eq. \eqref{eta1_} gives $\tilde{\Delta}=0$, i.e. for the original detuning $\Delta$ from \eqref{DefTD}:
 \begin{equation}
 \label{TDshape}
 \Delta= \Lambda_a - \Lambda_s \Bigl(\frac{1}{2}+\eta_{3}\Bigr).
 \end{equation} 
This leads to an optimal trajectory along the meridian connecting the south to the target near the north pole (of distance $\epsilon$ from it) in the $(\eta_{2}, \eta_{3})$ plane. The dynamics can be solved exactly from \eqref{etad_}. 
For instance, when we consider a population transfer from the ground state (south pole), i.e.  $\eta_{\text{3i}}=-1/2$, we obtain (taking $t_i=0$)
 \begin{equation}
 \label{soleta3Tmin}
 \eta_3(t)=\tanh^{2}\Bigl(\frac{1}{2}\Omega_0 t\Bigr)-\frac{1}{2}.
\end{equation} 
\begin{figure}
		\includegraphics[scale=0.75]{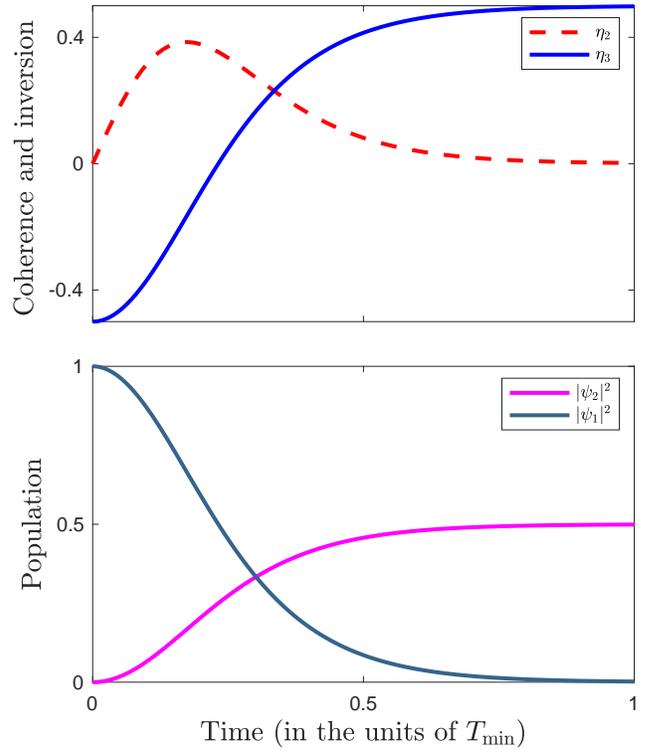}
		\caption{Populations (lower frame), coherence and population inversion (upper frame) history, given by \eqref{soleta3Tmin} and integration of \eqref{eta2_}, governed by the time optimal (constant) pulse $\Omega_0$ for $\epsilon=0.002$, giving $\mathcal{A}\approx7.60$, i.e. $T_{\text{min}}\approx7.60/\Omega_{0}$. }
		\label{fig11}	
\end{figure}
From Eq. \eqref{Omega}, taken at initial time, we obtain
$\Omega_{0}=2/\lambda_{2\text{i}}$ (independently of the initial value of $\lambda_{3,\text{i}}$). 
To obtain the explicit expression of the minimum time $T_{\min}=\min (t_f-t_i)$ for a given $\Omega_0$, we calculate the corresponding minimum pulse area by integrating \eqref{eta3_} using \eqref{surface}, which is fully determined by the (given) initial and final boundaries of $\eta_{3}$,
\begin{equation}
	\label{T}
		\mathcal{A}_{\min}\equiv \Omega_{0} T_{\min}=\pm\int_{\eta_{\text{3i}}}^{\eta_{\text{3f}}}\frac{d\eta_{3}}{\sqrt{\left(\frac{1}{2}-\eta_{3}\right)^{2}\left(\frac{1}{2}+\eta_{3}\right)}}.
\end{equation}
The $\pm$ sign ensures a non-negative pulse area, i.e. the sign $+$ ($-$) corresponds to $\eta_{\text{3i}}<\eta_{\text{3f}}$ ($\eta_{\text{3i}}>\eta_{\text{3f}}$). Hereby, we consider $\eta_{3\text{i}}<\eta_{3\text{f}}=\frac{1}{2}-\epsilon$ and finally get the minimum time for given $\epsilon$ and $\Omega_{0}$,
\begin{align}
T_{\text{min}}=\frac{2}{\Omega_{0}}\left|\text{atanh}\sqrt{\frac{1}{2}+\eta_{\text{3f}}}-\text{atanh}\sqrt{\frac{1}{2}+\eta_{\text{3i}}}\right|.
\label{Tmin}
\end{align}
This gives for the non-linear final transfer probability from the ground state in optimal time $T_{\min}$:
	\begin{equation}
	\label{NonlinearProba}
	p=2\vert\psi_2(T_{\min})\vert^2=\tanh^{2}\Bigl(\frac{1}{2}\Omega_0 T_{\min}\Bigr)=1-\epsilon.
		\end{equation}
We notice in the limit case of unbounded pulse amplitude a Dirac $\delta$ pulse, i.e. of infinite amplitude and zero duration with a finite area $\Omega_0T_{\text{min}}$ given by \eqref{Tmin}.
The dynamics from the ground state is shown in Fig. \ref{fig11} for $\epsilon=0.002$.

The minimum time $T_{\text{min}}$ can be used as the definition of the so-called quantum speed limit in this system as suggested in \cite{Frey}.
In linear systems, the mimum time is given by $T_{\min,\text{lin.}}$:  $\cos\bigl(\Omega_0T_{\min,\text{lin.}}/2\bigr)=\frac{1}{2}\sqrt{1-\eta_{3\text{i}}}\sqrt{1-\eta_{3\text{f}}} + \frac{1}{2}\sqrt{1+\eta_{3\text{i}}}\sqrt{1+\eta_{3\text{f}}}$ \cite{Boscain,Frey} with $\eta_{3\text{i}}=-1$ for the ground state and $\eta_{3\text{f}}=1$ for the excited state. In the non-linear case we obtain $\tanh(\Omega_{0} T_{\text{min}}/2)=\sqrt{\frac{1}{2}+\eta_{\text{3f}}}$ for $\eta_{3\text{i}}=-1/2$ from Eq. \eqref{Tmin}. 

\begin{figure}
	\begin{center}
		\includegraphics[scale=0.75]{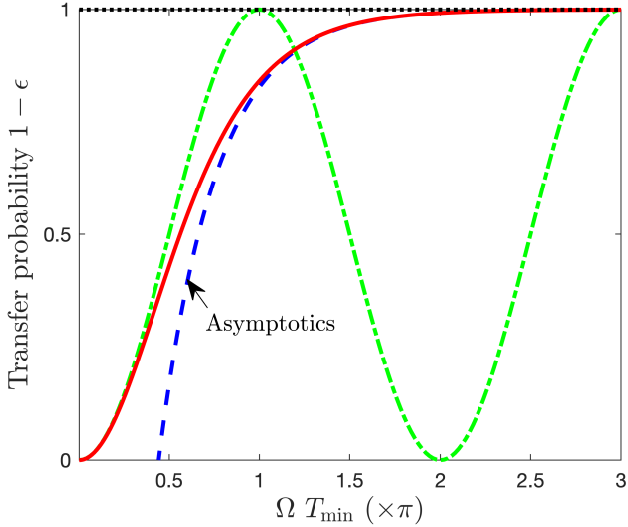}
		\caption{Population transfer probability $1-\epsilon$ as a function of the optimal pulse area for the linear $p_{\text{lin}}=\sin^{2}\left(\Omega_{0}T_{\min}/2\right)$ (green dotted-dashed line) and nonlinear (red full line) two-level models, respectively. Asymptotics \eqref{exponent} of the probability (blue dashed line) shows its accuracy when $\epsilon\to0$. 
		}
		\label{figrobust}
	\end{center}
\end{figure}

We consider $\eta_{\text{3i}}=-1/2$ (south pole). The pulse area $\mathcal{A}_{\min}=\Omega_{0}T_{\min}$ is given from Eq. \eqref{Tmin} with its leading order for $\epsilon\to0$:
\begin{equation}
	\label{exponent}
	\mathcal{A}_{\min}=2\, \text{atanh} \sqrt{1-\epsilon}\sim-\ln\Bigl(\frac{\epsilon}{4}\Bigr),\text{ i.e. } \epsilon\sim4e^{-\mathcal{A}_{\min}}.
	\end{equation}
We conclude that the minimum time (or pulse area) for the (almost complete) inversion increases in a (slow) logarithmic way with respect to small deviations $\epsilon$, or reciprocally that the deviation $\epsilon$ decreases exponentially with respect to the minimum time (or pulse area).
Figure \ref{figrobust} compares the pulse area cost between the linear, $p_{\text{lin}}=\sin^{2}\left(\Omega_{0}T_{\min}/2\right)$, and the non-linear probability \eqref{NonlinearProba}. One can observe that the non-linearity affects weakly the transfer for small transfers ($\epsilon\to1$).

\subsection{Energy optimal control}
We consider the cost functional for the energy optimal control 
\begin{equation}
	\label{energycost}
J\equiv E=\hbar\int^{t_{f}}_{t_{i}}\Omega^{2}(t)dt.
\end{equation}
In this case, we can rewrite the control Hamiltonian (in units such that $\hbar=1$)
\begin{align}
		h_{c}&=\tilde{\Delta}(\lambda_{1}\eta_{2}-\lambda_{2}\eta_{1})\nonumber\\
		& +\Omega\Bigl[\frac{\lambda_{2}}{2}\Bigl(3\eta^{2}_{3}-\eta_{3}-\frac{1}{4}\Bigr)+\lambda_{3}\eta_{2}\Bigr] -p_{0}\Omega^{2}
\end{align}
with the standard choice $p_0=1/2$. In this case, the costate has the angular frequency unit.
After applying the PMP, 
\begin{align}
\frac{\partial h_{c}}{\partial\tilde{\Delta}}=0,\quad \frac{\partial h_{c}}{\partial\Omega}=0,
\label{energypmptwo}
\end{align}
we obtain $\lambda_{1}\eta_{2}-\lambda_{2}\eta_{1}=0$ and
\begin{align}
	\label{Omegaenergy}
	&\Omega=\frac{\lambda_{2}}{2}\Bigl(3\eta^{2}_{3}-\eta_{3}-\frac{1}{4}\Bigr)+\lambda_{3}\eta_{2}.
\end{align}
The control Hamiltonian can be rewritten as 
\begin{align}
	h_{c}=\frac{1}{2}\Omega^{2},
\end{align}
leading to a constant coupling $\Omega=\Omega_0=\sqrt{2h_c}$. 
Similarly as for the time-optimum case, we obtain $\eta_1=0$, $\lambda_1=0$, $\tilde{\Delta}=0$, and thus the same dynamics as in Fig. 1. We also derive Eq. \eqref{Tmin} but interpreted differently, i.e. for a given time of interaction $T=t_{f}-t_{i}$, we determine the minimum $\Omega_0$:
\begin{align}
\Omega_{0,\min}=\frac{2}{T}\left|\text{atanh}\sqrt{\frac{1}{2}+\eta_{\text{3f}}}-\text{atanh}\sqrt{\frac{1}{2}+\eta_{\text{3i}}}\right|
\label{Omega0}
\end{align}
leading to the minimum energy 
\begin{align}
E_{\min}=\hbar\Omega_{0,\min}^{2}T=\hbar\mathcal{A}_{\min}^{2}/T
\end{align}
corresponding to the minimum area given by \eqref{Omega0}: $\mathcal{A}_{\min}=\Omega_{0,\min} T=2\left|\text{atanh}\sqrt{\frac{1}{2}+\eta_{\text{3f}}}-\text{atanh}\sqrt{\frac{1}{2}+\eta_{\text{3i}}}\right|$.

\section{1:2 nonlinear three-level $\Lambda$ model}
\label{model}

\subsection{The model}
The equations of motion for the three-level Raman model (forming a $\Lambda$ system), including the second- and third-order nonlinearities, read 
\begin{subequations}
	\label{H_}
	\begin{align}
		&i \dot{\psi}_{1}=K_1\psi_1+\Omega_{p}\psi^{*}_{1}\psi_{2},\\
		&i\dot{\psi}_{2}=K_2\psi_2+ \Delta_P\psi_{2}+\frac{\Omega_{p}}{2}\psi_{1}^{2}+\frac{\Omega_{s}}{2}\psi_{3},\\
		&i\dot{\psi}_{3}= K_3\psi_3+  \frac{\Omega_{s}}{2}\psi_{2} + (\Delta_P-\Delta_S)\psi_{3},
	\end{align}
\end{subequations}
where $\Omega_{p}$ and $\Omega_{s}$ are the time-dependent pump and Stokes fields, respectively,
$\Delta_P$ is the one-photon detuning associated to the pump coupling, and $\Delta_P-\Delta_S$ is the two-photon detuning associated to the Raman process.  The second-order 1:2 nonlinearity appears here through the pump coupling. This is typically the situation for the two-color photoassociation process \cite{pra65_Drummond}. The third-order nonlinearities $K_j$, $j=1,2,3$, can be absorbed in the definition of the detunings and the change of phases $\psi_{1}\to\psi_1e^{-i\gamma}$, $\psi_{2,3}\to\psi_{2,3}e^{-2i\gamma}$: $\dot\gamma=K_1$, $\Delta_P=2K_1-K_2$, $\Delta_S=K_3-K_2$, in order to lock the (one- and two-photon) resonances  \cite{prl119_Stephane}:
\begin{subequations}
	\label{H}
	\begin{align}
		&i \dot{\psi}_{1}=\Omega_{p}\psi^{*}_{1}\psi_{2},\\
		&i\dot{\psi}_{2}= \frac{\Omega_{p}}{2}\psi_{1}^{2}+\frac{\Omega_{s}}{2}\psi_{3},\\
		&i\dot{\psi}_{3}=  \frac{\Omega_{s}}{2}\psi_{2}.
	\end{align}
\end{subequations}
The amplitude probabilities satisfy $|\psi_{1}|^{2}+2(|\psi_{2}|^{2}+|\psi_{3}|^{2})=1$.
We decompose the components into real and imaginary parts, $\psi_{i}=x_{i}+iy_{i}$ $(i=1, 2, 3)$, and assume real Rabi frequencies, which allows one to separate the original problem into two disjoint dynamics that emerge according to the real and imaginary parts of the initial state, respectively.
We consider the initial state $\psi_{1}$ being real, $x_{1}(t_{i})=1$, and we get the equations of motion from \eqref{H}:
\begin{subequations}
	\label{xcoordinate}
	\begin{align}
		&\dot{x}_{1}=\Omega_{p}x_{1}y_{2},\\
		&\dot{y}_{2}=-\frac{\Omega_{s}}{2}x_{3}-\frac{\Omega_{p}}{2}x^{2}_{1},\\
		&\dot{x}_{3}=\frac{\Omega_{s}}{2}y_{2}.
	\end{align}
\end{subequations}
This system of equations can be analyzed using an isomorphism with the non-linear two-level problem, similarly to the linear problem \cite{ShoreVitanov} (see Appendix A). One can also show the incomplete transfer between the two ground states for finite pulse areas. We however prefer to keep the original coordinates for solving the problem.

Without loss of generality, we can parameterize the dynamics with the two dynamical angles $\theta(t)\in[0, \pi]$ and $\phi(t)\in[0, 2\pi[$:
\begin{subequations}
	\label{pangle}
	\begin{align}
		&x_{1}=\cos\phi\cos\theta,\\
		&y_{2}=-\frac{1}{\sqrt{2}}\sin\phi,\\
		&x_{3}=-\frac{\cos\phi\sin\theta}{\sqrt{2}},
	\end{align}
\end{subequations}
which satisfies the normalization condition $x_{1}^{2}+2(y_{2}^2+x_{3}^2)=1$. Inserting the definition \eqref{pangle} into Eqs. \eqref{xcoordinate} leads to
\begin{subequations}
	\label{angles}
	\begin{align}
		&\dot{\phi}=\frac{\Omega_{p}\cos\phi\cos^{2}\theta}{\sqrt{2}}-\frac{\Omega_{s}\sin\theta}{2},\\
		&\dot{\theta}=\frac{\sin\phi\sin\theta}{\cos\phi\cos\theta}\left(\frac{\Omega_{s}}{2\sin\theta}+\frac{\Omega_{p}\cos\phi\cos^{2}\theta}{\sqrt{2}}-\frac{\Omega_{s}\sin\theta}{2}\right),
	\end{align}
\end{subequations}
which by inversion provide the shape of the fields as functions of the angles:
\begin{subequations}
\label{pulsesPS}
	\begin{align}
		&\Omega_{s}=2\left(\dot{\theta}\cot\phi\cos\theta-\dot{\phi}\sin\theta\right),\\
		&\Omega_{p}=\sqrt{2}\left(\frac{\dot{\phi}}{\cos\phi}+\frac{\dot{\theta}\tan\theta}{\sin\phi}\right).
	\end{align}
\end{subequations}
We define the target state in the vicinity of the state 3:
	\begin{align}
	|x_{3}(t_{f})|^{2}=\frac{1}{2}(1-\epsilon),
		\end{align}
		where $\epsilon$ is, as in the two-level case, a small deviation due to the incomplete population transfer in such nonlinear system \cite{prl119_Stephane}.
To determine the expression of the optimal trajectories from the ground state to the target state (for a given $\epsilon$), we apply the PMP taking for the cost the time or the energy.

\subsection{Time optimal control}
\subsubsection{Definition}
We consider the time minimizing cost functional \eqref{MinTimeCost}
and impose similarly to the two-level case the constraint of bounded pulses amplitudes:
\begin{equation}
	\Omega^{2}_{p}+\Omega^{2}_{s}\le\Omega^{2}_{0}.
	\label{u1}
\end{equation}
The control Hamiltonian reads (where we have added a constant $p_{0}$)
\begin{align}
	\label{Htcontrol}
		H_{c}&=\lambda_{\phi}\dot{\phi}+\lambda_{\theta}\dot{\theta},\\
		&=\lambda_{\phi}\left(\frac{\Omega_{p}\cos\phi\cos^{2}\theta}{\sqrt{2}}-\frac{\Omega_{s}\sin\theta}{2}\right)\nonumber\\ 
		&+\lambda_{\theta}\left(\frac{\Omega_{s}\cos\theta\tan\phi}{2}+\frac{\Omega_{p}\cos\theta\sin\theta\sin\phi}{\sqrt{2}}\right),
\end{align}
where $\lambda_{\phi,\theta}$ are the component of the (dimensionless) costate $\Lambda=[\lambda_{\phi},\lambda_{\theta}]^{\mathsf{T}}$ of dynamics 
\begin{subequations}
	\label{adjoint}
\begin{align}
	&\dot\lambda_{\phi}=-\frac{\partial H_{c}}{\partial\phi}\nonumber\\
	&=\frac{\lambda_{\phi}{\Omega}_{p}\sin\phi\cos^{2}\theta}{\sqrt{2}}-\lambda_{\theta}\left(\frac{{\Omega}_{s}\cos\theta}{2\cos^{2}\phi}+\frac{{\Omega}_{p}\sin2\theta\cos\phi}{2\sqrt{2}}\right),\\
	&\dot{\lambda_{\theta}}=-\frac{\partial H_{c}}{\partial\theta}\nonumber\\
	&=\lambda_{\phi}\left(\frac{{\Omega}_{p}\cos\phi\sin2\theta}{\sqrt{2}}+\frac{\Omega_{s}\cos\theta}{2}\right)\nonumber\\
	&+\lambda_{\theta}\left(\frac{{\Omega}_{s}\sin\theta\tan\phi}{2}-\frac{{\Omega}_{p}\cos2\theta\sin\phi}{\sqrt{2}}\right).
\end{align}
\end{subequations}
Without loss of the generality, we can consider the control functions satisfying the contraint \eqref{u1} as follows:
	\begin{align}
	\label{controlf}
		\Omega_{p}=\Omega_m(t)\cos\beta(t),\quad
		\Omega_{s}=\Omega_m(t)\sin\beta(t)
	\end{align}
with the condition
\begin{equation}
	\Omega^{2}_{s}+\Omega^{2}_{p}=\Omega_m^2\le\Omega_0^2.
	\label{u2}
\end{equation}
The constraint \eqref{u1} is thus transferred to the condition \eqref{u2} (which is independent on $\beta$). The PMP maximization of $H_{c}$ is thus reduced to the necessary condition
\begin{equation}
	\frac{\partial H_{c}}{\partial \beta}=0,
\end{equation}
which gives
\begin{equation}
	\frac{\partial H_{c}}{\partial\beta}=\frac{\partial \Omega_{p}}{\partial\beta}H_{1}+\frac{\partial \Omega_{s}}{\partial\beta}H_{2}=-H_{1}\sin\beta+H_{2}\cos\beta=0,
\end{equation}
where
\begin{subequations}
\label{H1H2}
\begin{align} 
&H_{1}=\frac{\lambda_{\phi}\cos\phi\cos^{2}\theta}{\sqrt{2}}+\frac{\lambda_{\theta}\sin2\theta\sin\phi}{2\sqrt{2}},\\ &H_{2}=\frac{\lambda_{\theta}\cos\theta\tan\phi}{2}-\frac{\lambda_{\phi}\sin\theta}{2}.  
\end{align}
\end{subequations}
We then deduce
	\begin{align}
	\label{theta}
		\cos\beta=\frac{H_{1}}{\sqrt{H^{2}_{1}+H^{2}_{2}}},\quad \sin\beta=\frac{H_{2}}{\sqrt{H^{2}_{1}+H^{2}_{2}}}.
	\end{align}
Substituting Eq. \eqref{theta} in Eq. \eqref{Htcontrol}, we obtain
\begin{equation}
	H_{c}=\Omega_m\sqrt{H^{2}_{1}+H^{2}_{2}}.
\end{equation}
from which we conclude that the control Hamiltonian is maximum for $\Omega_m=\Omega_0$, i.e. when the maximum of \eqref{u1} is reached at all times.

\begin{figure}
	\centering
\begin{center}
	\includegraphics[scale=0.78]{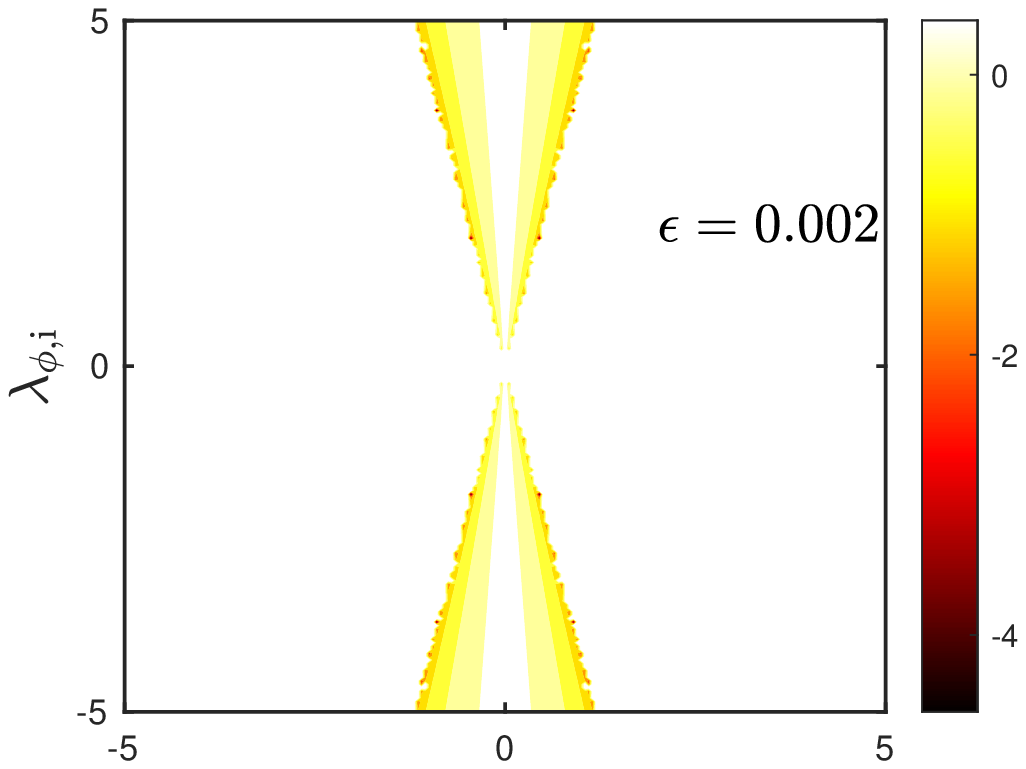}\\
		\includegraphics[scale=0.78]{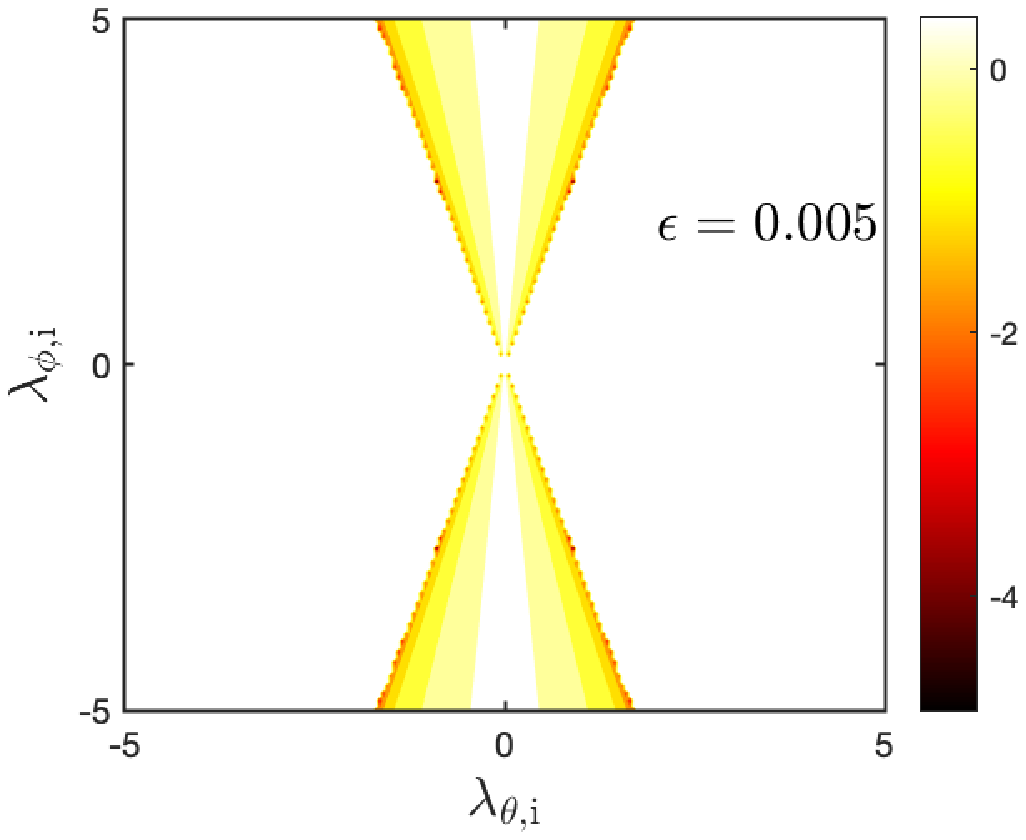}
		\caption{Contour plot for $\log_{10}(T-T_{\min})$ as a function of dimensionless $\lambda_{\phi,\text{i}}$ and $\lambda_{\theta,\text{i}}$ when the dynamics reaches the targeted transfer, for a given $\epsilon$. The transfer time is determined in a given range $T\in[0, \hat T]$ (taken here as $\hat T=15/\Omega_{0}$). The white areas mean that there is no transfer time found in the prescribed interval. We obtain $T_{\min}\approx7.40/\Omega_{0}$ for $\epsilon=0.002$ (upper frame), $T_{\min}\approx6.78/\Omega_{0}$ for $\epsilon=0.005$ (lower frame) along the dark straight lines, respectively.}
		\label{fig3}
	\end{center}
\end{figure}

\subsubsection{Numerics}
We first determine the systematic landscape of the time $T$ to reach the target (for a given $\epsilon$) by solving the set of equations (\ref{angles},\ref{adjoint}) with the controls \eqref{controlf} and $\Omega_m=\Omega_0$ rewritten as functions of the angles and the costate components via (\ref{theta},\ref{H1H2}), as a function of the two parameters $\lambda_{\phi,\text{i}}$ and $\lambda_{\theta,\text{i}}$, which with $\theta_{\text{i}}=0$, $\phi_{\text{i}}=0$ form the set of initial conditions.
The landscape is shown in Fig. \ref{fig3}, where the white areas corresponds to the absence of solution reaching the given target in the prescribed interval.
Figure \ref{fig3} shows an infinite set of initial $\lambda_{\phi\text{i}}$, $\lambda_{\theta\text{i}}$ forming two straight (symmetric) lines that lead to the same minimum $T_{\min}\approx7.4/\Omega_{0}$. 

The four quadrants give all the possible respective signs of the controls. The controls are both positive when the initial values $\lambda_{\phi\text{i}}$, $\lambda_{\theta\text{i}}$ are taken positive.
In order to determine a more accurate value of the optimum, we choose a certain value $\lambda_{\phi\text{i}}$ (\textit{e.g.} $\lambda_{\phi\text{i}}=1.85$) and run an optimal procedure leading to the (positive) value $\lambda_{\theta\text{i}}\approx0.45266$ corresponding to the minimum time (using a Nelder-Mead simplex algorithm as described in \cite{NMSM_matlab}).

\begin{figure}
	\begin{center}
		\includegraphics[scale=0.7]{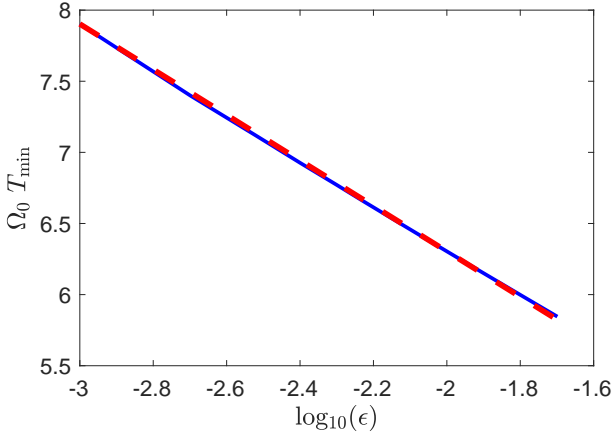}
		\caption{Optimal pulse area (blue full line) as a function of $\epsilon$ (in a logarithmic scale) for $\lambda_{\phi\text{i}}=1.85$ and the resulting optimum $\lambda_{\theta\text{i}}$
		and the logarithmic fit of the asymptotics for small $\epsilon$:  ${\cal A}_{\min}=\Omega_{0}T_{\min}\sim-(\ln\epsilon)/\sqrt{2}+3$ (red dashed line). 
		}
		\label{fig4}
	\end{center}
\end{figure}
\begin{figure}
		\begin{center}
			\includegraphics[scale=0.7]{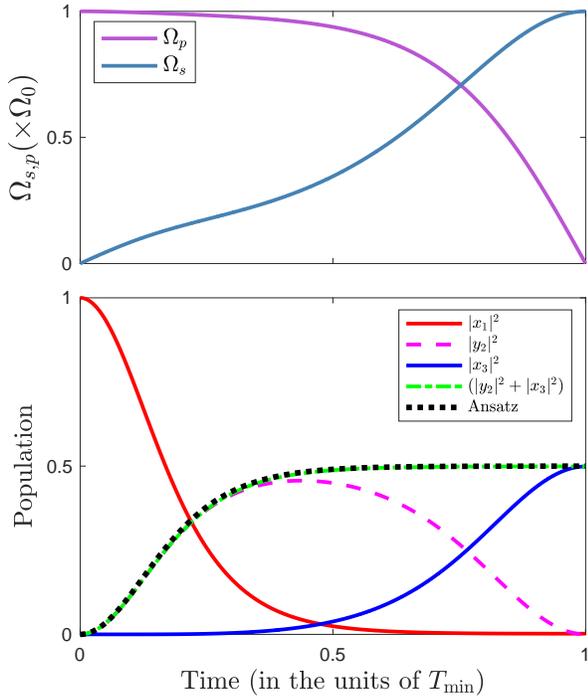}
			\caption{Optimal pulse shapes \eqref{pulsesPS} in units of $\Omega_0$ with $\epsilon=0.002$, $\lambda_{\phi\text{i}}=1.85$ and $\lambda_{\theta\text{i}}\approx0.45266$ operating in minimum time $T_{\min}\approx7.40/\Omega_{0}$ (upper frame), and the resulting population history and the ansatz \eqref{p10} (lower frame). }
			\label{fig5}
		\end{center}
\end{figure}

Figure. \ref{fig4} shows as a function of $\epsilon$ the minimum time, via the minimum generalized pulse area 
defined as
${\cal A}_{\min}=\int_{0}^{T_{\min}}dt \sqrt{\Omega_p^2+\Omega_s^2}=T_{\min}\Omega_0$. 
It exhibits a logarithmic decreasing behavior similar to the two-level case. 
This suggests the following ansatz for the population inspired by the two-level problem and the behavior of the optimal pulse area in Fig. \ref{fig4} (taking $t_i=0$):
\begin{align}
	y_2^2+x_3^2\approx\frac{1}{2}\tanh^{2}\Big(\frac{\Omega_{0}}{\sqrt{2}}t\Bigr),
	\label{p10}
\end{align}
which fits well the numerics of the dynamics shown in Fig. \ref{fig5}, but slightly overestimates the accuracy (of the order of $\epsilon$).
The boundary at $t_f=T_{\min}$ gives the asymptotic expansion for small $\epsilon$ (with $y_{2\text{f}}^2\ll x_{3\text{f}}^2$):
\begin{align}
	y_{2\text{f}}^2+x_{3\text{f}}^2 \approx \frac{1}{2}(1-\epsilon) \sim \frac{1}{2}\bigl(1-4e^{-\sqrt{2}{\cal A}_{\min}} \bigr),
\end{align}
i.e.
\begin{align}
	{\cal A}_{\min} \sim  -\frac{1}{\sqrt{2}}(\ln\epsilon-\ln4). 
\end{align}
We notice that the resulting logarithmic scaling with respect to $\epsilon$ is the same as the scaling of the fit determined from Fig. \ref{fig4}. The absolute value is different due to the systematic error of \eqref{p10} mentioned above.

\begin{figure}
		\begin{center}
			\includegraphics[scale=0.7]{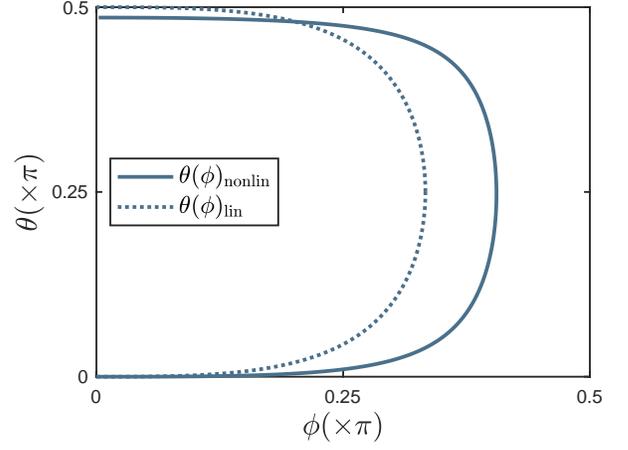}
			\caption{Trajectories in the $\phi,\theta$ space of the optimal non-linear solution corresponding to the dynamics shown in Fig. \ref{fig5} (full line) and the optimal solution of the linear model (dotted line).}
			\label{fig5_}
		\end{center}
\end{figure}

The dynamics and the controls of Fig. \ref{fig5} show an intuitive sequence of pulses with a large transient population in the upper state, similarly to the linear case which features pulses of explicit form $\cos -\sin$ \cite{Boscain,TNXu}. Figure  \ref{fig5_} shows the trajectory in the angles $\phi,\theta$ space corresponding to the optimal non-linear solution for $\epsilon=0.002$, it is compared to the optimal solution of the linear problem. 

\subsection{Energy optimal control}
For the energy optimization, we use the cost functional
\begin{equation}
	\label{energycost1}
	J\equiv E=\hbar\int^{t_{f}}_{t_{i}}(\Omega^{2}_{p}+\Omega^{2}_{s})dt.
\end{equation}
We derive the control Hamiltonian (using the standard value $p_0=1/2$):
\begin{align}
	H_{c}&=\lambda_{\phi}\dot{\phi}+\lambda_{\theta}\dot{\theta}-p_{0}(\Omega_{p}^{2}+\Omega_{s}^{2}),
	\nonumber\\
		&=\lambda_{\phi}\left(\frac{\Omega_{p}\cos\phi\cos^{2}\theta}{\sqrt{2}}-\frac{\Omega_{s}\sin\theta}{2}\right)\nonumber\\ 
		&+\lambda_{\theta}\left(\frac{\Omega_{s}\cos\theta\tan\phi}{2}+\frac{\Omega_{p}\cos\theta\sin\theta\sin\phi}{\sqrt{2}}\right)\nonumber\\
	    &-\frac{1}{2}(\Omega_{p}^{2}+\Omega_{s}^{2}).
\end{align}
The PMP $\frac{\partial H_{c}}{\partial\Omega_p}=0$, $\frac{\partial H_{c}}{\partial\Omega_s}=0$ leads to the pulse shape:
\begin{subequations}
	\label{pulses1}
	\begin{align}
		&\Omega_{p}=\frac{\lambda_{\phi}\cos\phi\cos^{2}\theta}{\sqrt{2}}+\frac{\lambda_{\theta}\sin2\theta\sin\phi}{2\sqrt{2}},\\
		&\Omega_{s}=\frac{\lambda_{\theta}\cos\theta\tan\phi}{2}-\frac{\lambda_{\phi}\sin\theta}{2},
	\end{align}
\end{subequations}
and 
\begin{equation}
	H_{c}=\frac{1}{2}({\Omega^{2}_{p}+\Omega^{2}_{s}}),
\end{equation}
which is constant as before, i.e. $\Omega^{2}_{p}+\Omega^{2}_{s}=\Omega_0^2$.
The dynamics of the components of the (angular frequency unit) costate $\Lambda=[\lambda_{\phi},\lambda_{\theta}]^{\mathsf{T}}$ is still given by \eqref{adjoint}.
We thus obtain the same dynamics shown in Fig. \ref{fig5} as in the time-optimal control but for a given time of interaction $T=t_{f}-t_{i}$ (instead of $T_{\min}$). The corresponding optimal generalized pulse area ${\cal A}_{\min}$ is determined from Fig. \ref{fig4} for a given $\epsilon$: ${\cal A}_{\min}\approx-\ln(\epsilon)/\sqrt{2}+3$, and we deduce the corresponding $\Omega_{0,\min}={\cal A}_{\min}/T$. The minimum energy 
is then
\begin{align}
E_{\min}=\hbar\Omega_{0,\min}^2T=\hbar\mathcal{A}_{\min}^{2}/T.
\end{align}

\section{Conclusion}
\label{conclusion}

In this paper, we have determined the ultimate bounds in terms of optimal time and optimal energy for the two- and Raman three-level problems featuring a 1:2 non-linear resonance when an accurate (but not strictly complete) population transfer is targeted.
In both cases, we have incorporated the third-order Kerr terms in the detuning locking the dynamics to the resonance at all times.
In the two-level system, we have shown the equivalence of the dynamics for the optimal time or energy, given by a resonant and constant pulse, as it the case for the linear problem. The behavior of the resonant nonlinear dynamics is qualitatively different with the linear one: the complete inversion in only (exponentially) asymptotic instead of the Rabi oscillations of the linear problem (see Fig. \ref{figrobust}). The optimal time features an asymptotic logarithmic increasing as a function of the accuracy.
For the three-level problem, the optimal solution can be obtained only numerically. However, we have fitted it using the resembling results of the two-level problem. In this case, the generalized pulse area is constant. We have determined the shape of the individual pulses featuring an intuitive pump-Stokes sequence, as in the linear case but with different shapes.
We have also obtained an asymptotic logarithmic increasing of the optimal time as a function of the accuracy.

The finding of the ultimate bounds (time or energy) for nonlinear systems provide an important benchmark.
The issue of robustness of the process will have to be considered in future analysis involving optimal inverse engineering \cite{Dridi}, since the (almost) complete transfer is very unstable when the resonance is not perfectly satisfied (see Fig. 1 of Ref. \cite{pra_SG}). This is also the case for two-level adiabatic transfer \cite{pra102_Jingjun}. We notice that this instability does not exist when only third-order nonlinearities apply, neither for the usual nonlinear Raman three-level processes, where the target state is linked with the Stokes coupling which is linear.

\begin{acknowledgments}
We acknowledge support from the EUR-EIPHI Graduate School (17-EURE-0002) and from the European Union's Horizon 2020 research and innovation program under the Marie Sklodowska-Curie Grant No. 765075 (LIMQUET).  
X.C. acknowledges  EU FET Open Grant EPIQUS (Grant No. 899368), the
Basque Government through Grant No. IT1470-22,
the project grant PID2021-126273NB-I00 funded by
MCIN/AEI/10.13039/501100011033 and by “ERDF A way
of making Europe” and “ERDF Invest in your Future” and ayudas para contratos Ramon y Cajal–2015-2020 (RYC-2017-22482).

\end{acknowledgments}

\appendix
\section{Isomorphism between non-linear three-level and two-level models}
The three-level problem \eqref{xcoordinate} can be rewritten as
\begin{subequations}
\label{motion3level}
	\begin{align}
		&\dot{x}_{1}=-Px_{1}z_{2},\\
		& \dot{z}_{2}=Sz_{3}+Px^{2}_{1},\\
		&\dot{z}_{3}=-Sz_{2},
	\end{align}
\end{subequations}
with $z_2=-y_2\sqrt{2}$, $z_3=x_3\sqrt{2}$, $P=\Omega_{p}/\sqrt{2}$, $S=\Omega_{s}/\sqrt{2}$, and the normalization condition $x_{1}^{2}+z_{2}^2+z_{3}^2=1$.
We assume that $P$ and $S$ are real and that $x_1(t_i)=1$. It can be reinterpreted as a density ``matrix'' formulation on the Bloch sphere:
\begin{equation}
\frac{d}{dt}\left[\begin{array}{ccc} \rho_z \\ \rho_y \\ \rho_x \end{array}\right]=
\left[\begin{array}{ccc} 0 & -P \rho_z & 0 \\ P \rho_z
& 0 & S \\
0 & -S & 0 \end{array}\right]\left[\begin{array}{ccc} \rho_z \\ \rho_y \\ \rho_x \end{array}\right]
\end{equation}
with $\rho_z=x_1=\rho_{11}-\rho_{22}=\vert a_1\vert^2-\vert a_2\vert^2$, $\rho_y=z_2=i(\rho_{21}-\rho_{12})=2 \Im(a_1\bar a_2)$, $\rho_x=z_3=\rho_{21}+\rho_{12}=2 \Re(a_1\bar a_2)$, $\rho_{ij}=a_i \bar a_j$ of a non-linear two-level problem
\begin{equation}
\label{H2}
H_{2,n\ell}=\frac{1}{2}\left[\begin{array}{cc}-S & P(|a_1|^2-|a_2|^2)\\
P(|a_1|^2-|a_2|^2) & S\end{array}\right]
\end{equation}
with $i\frac{d}{dt} \left[a_1\ a_2\right]^t =H_{2,n\ell} \left[a_1\ a_2\right]^t $ and the normalization $|a_1|^2+|a_2|^2=1$.
As a consequence, the non-linear three-level problem \eqref{motion3level} is isomorphic to the above non-linear two-state problem \eqref{H2}.
The non-linearity which appears here is not the one usually encountered \eqref{motiontwolevel}. We obtain a similar isomorphic relation for linear problems \cite{ShoreVitanov}.

As a consequence, the transfer is, at the final time $t_f$, complete, $|z_3(t_f)|=1$, i.e. $\rho_x(t_f)=\pm1$, when, in the counterpart two-state problem, from the initial state $\rho_{11}(t_i)=1$, the superposition of state of maximal coherence is produced: $\rho_{12}(t_f)=\rho_{21}(t_f)=\pm1/2$.  This shows a similar qualitative behavior than for its linear analog.

The general solution of the two-state problem \eqref{H2} can be parametrized by three angles in general:
\begin{equation}
\label{solgen}
\left[\begin{array}{cc}a_1\\a_2\end{array}\right]
=\left[\begin{array}{cc}\cos(\theta/2)\\
\sin(\theta/2)e^{-i\varphi}\end{array}\right]e^{-i\gamma},
\end{equation}
and the Schr\"odinger equation leads to the set of equations

\begin{subequations}
	\label{Modelangles1}
	\begin{align}
		\label{ModelanglesI}
		&\dot \theta=P\cos\theta\sin\varphi,\\
		\label{Modelanglesal}
		&\dot\varphi=S+P\frac{\cos^2\theta}{\sin\theta}\cos\varphi,\\
		\label{Modelanglesgam}
		&\dot\gamma=-\frac{S}{2}+\frac{P\cos\theta\tan(\theta/2)\cos\varphi}{2}.
	\end{align}
\end{subequations}

One can solve Eq. \eqref{ModelanglesI} exactly [for any $P(t)$ and $S(t)$]:

\begin{equation}
		\label{solth}
		\tan(\theta/2)=\tanh\left[\frac{1}{2}\int_{t_i}^tP(s)\sin\varphi(s)ds \right].
\end{equation}

This shows that, in order to have a complete transfer from state 1 to state 3 in the original model, i.e. $\theta(t_i)=0$ and $\theta(t_f)=\pi/4$, one needs an infinite pulse area of $P(t)$. This contrasts with the linear model where a complete population transfer is possible for finite pulse areas \cite{Boscain}.

\end{document}